\documentclass[runningheads]{llncs}
\usepackage[T1]{fontenc}
\usepackage{graphicx}
\begin{document}
\title{Predicting the trajectory of intracranial pressure in patients with traumatic brain injury: evaluation of a foundation model for time series}
\titlerunning{Predicting the trajectory of intracranial pressure in patients with TBI}
%
\author{Florian D. van Leeuwen\inst{1}\orcidID{0009-0009-7092-2848} \and
Shubhayu Bhattacharyay\inst{2}\orcidID{0000-0001-7428-5588} \and
Alex Carriero \inst{3}\orcidID{0009-0007-4499-8043}
\and
Ethan Jacob Moyer\inst{4} \and
Richard Moberg\inst{4}}
\authorrunning{F. D. van Leeuwen et al.}
%
\institute{
Department of Methods and Statistics, Faculty of Social Science, Utrecht University, Utrecht, The Netherlands \\
\email{f.d.vanleeuwen@uu.nl}\\
\and
Harvard Medical School, 25 Shattuck St, Boston, MA 02115, USA \and
Julius Center for Health Sciences and Primary Care, University Medical Center Utrecht, Utrecht, The Netherlands
\and
Moberg Analytics, Inc, Philadelphia, PA, USA 
}
\maketitle              
\begin{abstract}
Patients with traumatic brain injury (TBI) often experience pathological increases in intracranial pressure (ICP), leading to intracranial hypertension—a common and serious complication. Early warning of an impending rise in ICP could potentially improve patient outcomes by enabling preemptive clinical intervention. However, the limited availability of patient data poses a challenge in developing reliable prediction models.
In this study, we aim to determine whether foundation models, which leverage transfer learning, may offer a promising solution. We compare a foundation model for time series (MOMENT) with established time-series architectures, including long short-term memory (LSTM) and exponential smoothing (ES), for forecasting ICP. We found that for forecasting ICP with a 30-minute horizon, the MOMENT and LSTM models yielded similar results and both outperformed the ES model.  However, all models performed poorly in predicting volatile periods, and there was substantial variability in model performance between patients. 

\end{abstract}


\section{Introduction}

Traumatic brain injury (TBI) afflicts millions of people annually. In Europe, it is estimated that each year 1 million individuals are admitted to the hospital, and 75,000 die as a result of TBI \cite{van_der_naalt_collaborative_2015}. TBI is a highly complex condition due to its heterogeneous nature: "TBI is considered the most complex disease in our most complex organ. It is characterized by great heterogeneity in terms of etiology, mechanisms, pathology, severity, and treatment, with widely varying outcomes" \cite[p.~68]{van_der_naalt_collaborative_2015}.

For moderate-to-severe TBI patients,\footnote{The initial clinical severity of TBI is often assessed using the Glasgow Coma Scale (GCS), which is commonly trichotomized into three classes of severity: mild (GCS 13-15), moderate (GCS 9-12), and severe (GCS $\leq$ 8) \cite{mild_traumatic_brain_injury_committee_definition_1993}, although the usefulness of this trichotomy is a point of discussion \cite{tenovuo_assessing_2021}.} A common complication is traumatic intracranial hypertension (tIH), which negatively affects patient outcomes \cite{stocchetti_traumatic_2014}. tIH occurs when the volume within the skull rapidly increases beyond the skull's capacity \cite{mcnamara_development_2023}, causing a spike in intracranial pressure (ICP). By monitoring ICP, potential tIH episodes can be detected. Clinicians may then decide whether to intervene, potentially exposing the patient to riskier treatments, or to tolerate the rise in ICP. While detecting a tIH event might prompt ad-hoc intervention, it remains unclear whether such interventions are beneficial after a damaging tIH episode \cite{cooper_decompressive_2011,hutchinson_trial_2016}. To provide the best care (or at least improved care), an early warning of a pending tIH episode would be beneficial for clinicians, potentially allowing for timely prevention of the episode \cite{mcnamara_development_2023}.

Such a warning can potentially be achieved through clinical prediction modeling. Many different models for ICP forecasting (e.g., \cite{ye_machine_2022,farhadi_intracranial_2019,han_online_2013}) and tIH prediction (e.g., \cite{carra_prediction_2021,schweingruber_recurrent_2022}) have been proposed. However, the process is not straightforward. In developing these models, decisions must be made regarding the features (variables used as inputs in the model), the length of the input sequence (how long to monitor a patient before making a prediction), the forecast horizon (how far into the future the forecast extends), and, in the case of tIH prediction, how to define a tIH event. Furthermore, tough choices regarding the model development (e.g., hyper parameters) need to be made with limited data, as the sample size is usually low. There is significant variation in these choices; for a comprehensive overview, refer to \cite{mcnamara_development_2023}. Although predicting a tIH event may be easier than forecasting the ICP signal, it might not provide enough clinical information to guide decision-making.

In this study, we focus on forecasting the ICP signal. The raw ICP signal is very noisy and requires preprocessing before it can be used in a model. Often, parts of the signal are not recorded (typically at the start or the end), there may be frequent interruptions or artifacts due to clinical management, or unrealistic values are measured. Approaches such as denoising and smoothing have been applied to ICP signals before the data is fed into a model for making predictions \cite{ye_machine_2022}.

There is an abundance of methodologies used to model time series, which are applied in many domains \cite{de_gooijer_25_2006}. Widely adopted algorithms, such as Exponential Smoothing (ES) \cite{holt_forecasting_1957} and ARIMA \cite{box_time_1970}, have been used for decades. There are also newer neural network-based sequence models, such as RNNs \cite{elman_finding_1990,hochreiter_long_1997} and transformers \cite{vaswani_attention_2017}. It is not always clear which model performs best in the healthcare domain \cite{kaushik_ai_2020}.

More recently, a foundation model for time series was published \cite{elman_finding_1990,hochreiter_long_1997}. A foundation model is a model trained on a broad dataset (usually through self-supervised learning) that can be fine-tuned for a wide range of tasks \cite{bommasani_opportunities_2022}. These models contain prior information about the task and/or domain for which they are designed. Foundation models have pre-trained weights that are adapted to new tasks using transfer learning \cite{thrun_lifelong_1995}. While transfer learning enables foundation models, it is the scale of these models (i.e., the number of parameters) that enables their true power \cite{bommasani_opportunities_2022}. The proposed MOMENT model is a 385-million-parameter model trained on 13 million unique time series from different domains (e.g., medical, energy, nature)\footnote{The datasets can be found here: https://huggingface.co/datasets/AutonLab/Timeseries-PILE} using self-supervised learning and tested on several different tasks.\footnote{Long-horizon forecasting, short-horizon forecasting, classification, anomaly detection, imputation.} In healthcare, where data is often limited, a foundation model may help improve predictions \cite{moor_foundation_2023}. In this study, we specifically investigate whether foundation models, and thus transfer learning, are helpful for ICP forecasting.

The main contributions of this study are as follows:

\begin{itemize} 
\item Outlining the preprocessing procedure for ICP signals, 
\item Comparing the performance of a foundation model (MOMENT) to established time-series model architectures that do not employ transfer learning (ES and LSTM) for ICP forecasting, 
\item Inspecting the performance of the models per patient and per forecast made for each patient. \end{itemize}

This paper is structured as follows: Section 2 outlines the data and methodology used in the analysis. Section 3 presents the results, and Section 4 provides the discussion.

\section{Methods}

\subsection{Data}
The training and internal validation data were from the high-resolution multimodal dataset from TRACK-TBI \footnote{More information can be found here: https://tracktbi.ucsf.edu}. TRACK-TBI was a prospective, multicenter observational study conducted at 18 Level 1 trauma centers in the US, enrolling patients with TBI between February 26, 2014, and August 8, 2018. The study was approved by the institutional review board of each TRACK-TBI site. Participants, or their legally authorized representatives, provided written informed consent to participate. The data were accessed through the Moberg AI Ecosystem (Moberg Analytics, Inc, Philadelphia, PA, USA). The data was obtained data through the FITBIR (Federal Interagency Traumatic Brain Injury Research) Informatics System under a data use agreement.

We have ICP data for multiple patients, collected from three different sites. Individual patients can have multiple recordings. We only used recordings with a duration of two hours or more, to ensure sufficient data for training and evaluation. We started with a total of 39 patients and 94 recordings. After removing unrealistic signals (see Appendix B for details), we ended up with 32 patients and a total of 83 recordings, comprising 5,142 hours of data. The distribution of recording times is shown in Figure \ref{fig:recording} (Appendix A). We only used data from patients whose ICP was measured with an intraparenchymal fiberoptic monitor (within the brain tissue). Data from patients with an external ventricular drain were not used, as their measurements are affected by interventional draining of cerebrospinal fluid. 

For external validation purposes, we use the CHARIS database \cite{kim_trending_2016}, which is publicly available\footnote{The data are available here: https://physionet.org/content/charisdb/1.0.0/} on PhysioNet \cite{goldberger_physiobank_2000}. The ICP data were monitored using either a subarachnoid bolt or ventriculostomy in 13 patients diagnosed with TBI. All patients were from the same hospital and had only one recording session. The data were preprocessed as described in Appendix B. To match the data used in \cite{ye_machine_2022}, parts of the recordings were removed based on the figures of the signals shown in \cite{ye_machine_2022}. For 10 of the 13 signals, the last part (10-120 hours) was removed. After preprocessing, there were a total of 1,122 hours of recordings.

ICP is measured in mm Hg; normal values for a person in a supine position range between 0.9 and 16.3 mm Hg \cite{pinto_increased_2024}. The Brain Trauma Foundation guidelines set the threshold for a tIH event at 22 mm Hg, though this is contested, and other values (e.g., 20 mm Hg) are sometimes used \cite{stocchetti_traumatic_2014,mcnamara_development_2023}.

\subsection{Preprocessing}
\begin{figure}[hbt!]
    \centering
    \includegraphics[width=.8\linewidth]{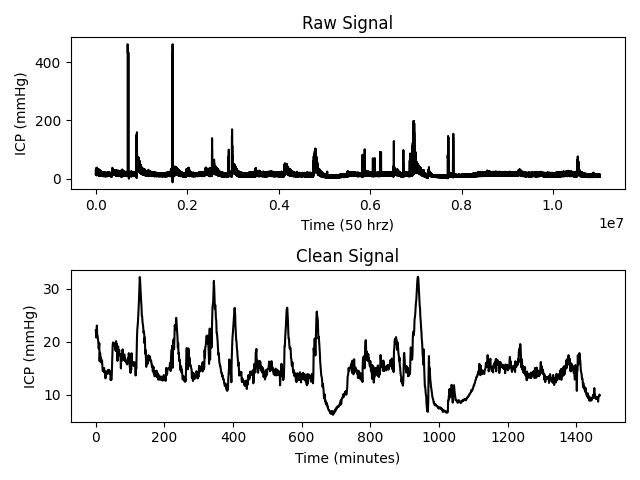}
    \caption{The raw versus pre-processed data for a random recording.}
    \label{fig:data_cleaning}
\end{figure}

ICP data are not "clean" (ready for model development); there is a significant amount of uncertainty (noise) in the signal. Some of this uncertainty arises from measurement error and missing observations \cite{gruber_sources_2023}. We want the algorithms to learn the noiseless ICP signal. To assist in this, we use the preprocessing step explained in Figure \ref{fig:pre_process} (Appendix B). The output of this preprocessing is a segment where each data point represents the ICP (mm Hg) value for a second. Figure \ref{fig:data_cleaning} provides an example of the transformation before and after preprocessing. The preprocessing and down-sampling appear to be effective.

Many of the signals have an unrealistic ending (Figure \ref{fig:pre_proc_bad_end}). It seems that the measuring device was not functioning correctly during the last part of some recordings, possibly due to the premature disconnection of the sensor. There are also signals for which the preprocessing did not eliminate the noise well (Figure \ref{fig:pre_proc_bad_full}); these were removed (11 out of 94).

After preprocessing, each minute has one ICP observation. We believe that a forecasting horizon of 30 minutes (similar to \cite{farhadi_intracranial_2019,guiza_early_2016,carra_prediction_2021}) provides clinicians with enough information to assist in the decision-making process. The input length is set to 60 minutes, meaning we use the previous 60 minutes of ICP data to forecast the next 30 minutes. To train and validate the model, we segment the data by cutting the signal into 90-minute sections (60 for X and 30 for Y). This process is further explained in Figure \ref{fig:segmentation} (Appendix B).

Both datasets were scaled by subtracting the mean and dividing by the standard deviation of all signals in the training set. 

\subsection{Models}

In this study, we compare three models: simple ES, a Recurrent Neural Network (LSTM), and the MOMENT model. The main characteristics of the models can be seen in Table \ref{tab:models}, and are explained further here. For all models, the only input variable was ICP. 

\begin{table}[htbp]
\centering
\caption{The characteristics of the models used. }
\label{tab:models}
\begin{tabular}{|c|c|c|c|c|}
\hline
\textbf{Model} & \textbf{Requires extensive training} & \textbf{Input-length} & \textbf{Output-length} & \textbf{Only Univariate} \\ \hline
MOMENT         & Yes (Fine-tuning)                & 512                   & Variable               & Partly $^a$                   \\
LSTM           & Yes                        & Variable              & Variable               & No                       \\
ES             & No                         & Variable              & Variable               & Yes                      \\ \hline
\end{tabular}
\vspace{2pt} 
\parbox{\linewidth}{\footnotesize $^a$ It can independently model univariate time series.}
\end{table}

The ES model is a classical statistical model that makes predictions without requiring a lengthy tuning procedure. It is a univariate model which can theoretically handle any input or forecast length. The model uses weighted averages of past observations, where more recent observations are given higher weights.

The Recurrent Neural Network (RNN) is a more flexible machine learning algorithm typically chosen for time series prediction tasks in healthcare \cite{morid_time_2023}. It can handle any input or output length and can handle multiple input features. In our case the RNN uses an LSTM cell with a size of 512. The architecture includes an encoder and a decoder, and teacher forcing (i.e., using the ground truth instead of the prediction) is applied 50\% of the time.

The MOMENT model\footnote{In the paper, three models were trained: small, base, and large. Their respective sizes are 40, 125, and 385 million parameters. Only the large model is publicly available and can be found here: https://huggingface.co/AutonLab/MOMENT-1-large. We use the large model in this study.} is a transformer with an encoder-decoder architecture. The MOMENT model is trained on various time series datasets by masking small parts of a time series and then reconstructing the masked parts. The model input is a univariate time series of length 512. The MOMENT model can handle multiple input features independently. The model can be adapted to any forecasting horizon by replacing the reconstruction head with a forecasting head. This relatively small prediction head maps the output of the encoder to the desired output length (65536 x output length). The prediction head needs to be fine-tuned for the task at hand while the parameters of the encoder are fixed. Dropout was included in the prediction head, set at 0.1 (default setting). See \cite{goswami_moment_2024} for a complete explanation of the MOMENT architecture.

We use a mean squared error (MSE) loss function with an Adam optimizer \cite{kingma_adam_2017} and a learning rate of 1e-5 for both the LSTM and the MOMENT models. A batch size of 64 was used for both the training and validation sets. Gradient clipping was applied with a maximum value of 5. \textit{PyTorch (2.4.0)} was used to implement the architectures for the MOMENT and LSTM models. Both were trained for 10 epochs. For the ES model, we used the \textit{tsa.holtwinters.ExponentialSmoothing} function from the \textit{statsmodels} package (0.14.2). All analyses were performed in \textit{Python} (3.10.12).

\subsection{Performance assessment}
The performance of the different models is assessed using MSE and mean absolute error (MAE), both commonly used for time series forecasting \cite{adhikari_introductory_2013}. There is a nested structure in the data, as each patient has multiple segments where the metrics are calculated, and there are multiple patients. For further explanation, see Appendix C.

We create a training and validation set by randomly sampling patients. We use 80\% of the patients for training and 20\% for validation. Some patients have multiple recordings. Due to the heterogeneous nature of TBI and the limited number of patients, splitting the training and validation data by patient ID can lead to significant variations in the performance metrics, depending on how patients are randomly assigned to each group. To ensure that we capture the performance of the models accurately, we perform k-fold cross-validation (CV) with k = 5. This will give us an indication of the internal validity of the model \cite{collins_evaluation_2024}. The number of recording samples per training and validation set were not uniform across CV partitions, due to the varying recording lengths and varying number of recordings per patient. The results presented for the internal validation will be the mean and the standard deviation (SD) of the performance measures over the 5-folds.

To assess the performance of the models in a new setting, we perform an external validation using the CHARIS database. Because CV does not yield one final model, we retrain the MOMENT and LSTM models on all available data from TRACK-TBI. The performance of all models is then assessed on the CHARIS dataset. All model parameters are held constant, based on the values used in the CV.

\section{Results}

\subsection{Internal validation}

Table \ref{tab:model_comparison_val} shows the results for the internal validation for each model. We see that the MOMENT model performs, based on the MSE and MAE, a little better than the LSTM, and both perform much better than the ES model in forecasting the next 30 ICP values (minutes) based on the previous 60 values. The lowest MAE was 1.78 (MOMENT). Although the MOMENT models outperformed LSTM and ES, the error is still considerably high given the normal range (without an iTH event) is between 0.9 and 16.3. For certain (more difficult) segments, all three models had equally poor performance. In the case of the MOMENT model, 10\% of the segments have a MAE value higher than 3.85 and 1\% higher than 9.45. The SD of the metrics is high for all models, which indicates heterogeneity in the data. The results for the training set can be found in Appendix D, Table \ref{tab:training_performance}. In the training set, the LSTM and MOMENT model perform similarly and again better than the ES model. The SD of the metrics is a lot lower in the training set. 


\begin{table}[htbp]
\caption{Average internal validation performance over 5 CV folds, SD is in brackets.}
\centering
\label{tab:model_comparison_val}
\begin{tabular}{|c|c|c|c|}
\hline
\textbf{Metric} & \textbf{MOMENT} & \textbf{LSTM} & \textbf{ES} \\
\hline
MSE & 9.06 (3.70) & 10.19 (3.50) & 22.56 (8.31) \\
MAE & 1.78 (0.40) & 1.86 (0.32) & 3.04 (0.61) \\
90th percentile MAE & 3.85 (0.78) & 3.91 (0.72) & 6.43 (1.31) \\
99th percentile MAE & 9.45 (1.98) & 11.13 (3.53) & 14.48 (1.57) \\
\hline
\end{tabular}
\end{table}

\subsubsection{Individual forecasts:}

\begin{figure}[h!]
  \centering
  \begin{minipage}[b]{0.49\textwidth}
    \includegraphics[width=\textwidth]{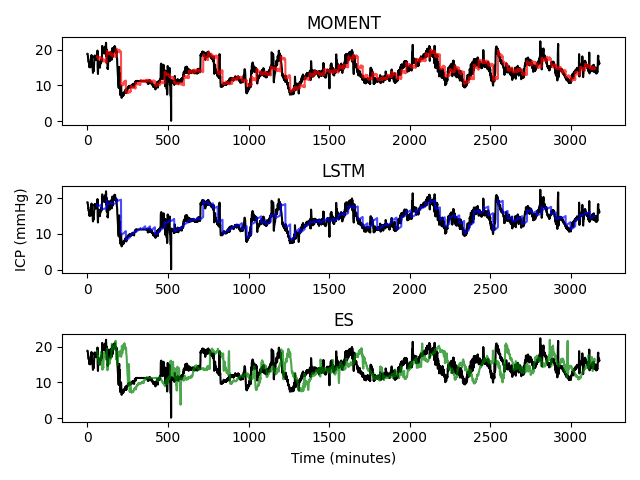}
    \caption{ICP signal with "good" forecast.}
    \label{fig:pred_patient_good}
  \end{minipage}
  \hfill
  \begin{minipage}[b]{0.49\textwidth}
    \includegraphics[width=\textwidth]{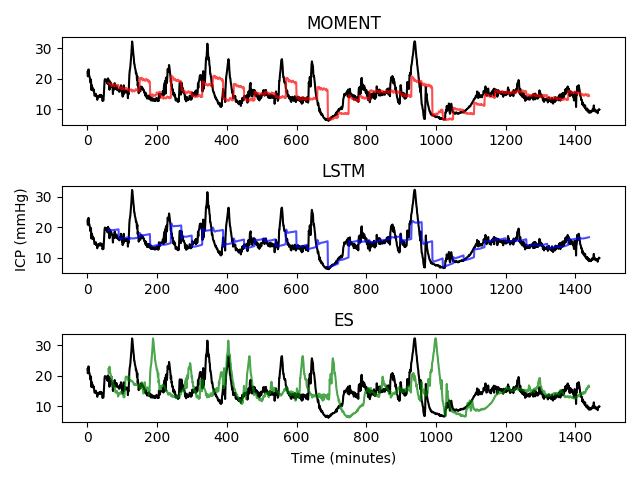}
    \caption{ICP signal with "bad" forecast.}
    \label{fig:pred_patient_bad}
  \end{minipage}
\end{figure}

In Figures \ref{fig:pred_patient_good} and \ref{fig:pred_patient_bad} we can see the (appended) 30-minute predictions with the observed time series (black) for two patients. The patient in Figures \ref{fig:pred_patient_good} has a more stable signal then the patient in Figures \ref{fig:pred_patient_bad}. We see that the predictions of the MOMENT (red) and LSTM (blue) models resemble the observed signal more closely than the ES (green) model. The models work quite well for the patient in Figure \ref{fig:pred_patient_good}, but are substantially worse for the patient in Figure \ref{fig:pred_patient_bad}. 

We will continue discussing the results of the MOMENT and LSTM models, as these performances are relatively close and much better than the ES model. 

\begin{figure}[h!]
  \centering
  \begin{minipage}[b]{0.49\textwidth}
    \includegraphics[width=\textwidth]{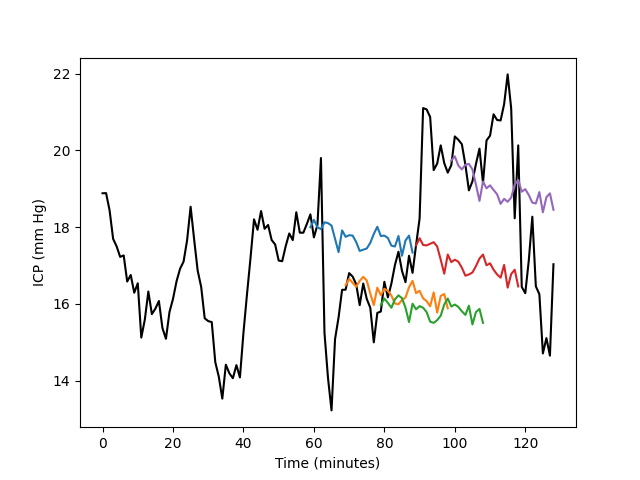}
    \caption{30-minute forecast every 10 min (MOMENT).}
    \label{fig:pred_5_min_moment}
  \end{minipage}
  \hfill
  \begin{minipage}[b]{0.49\textwidth}
    \includegraphics[width=\textwidth]{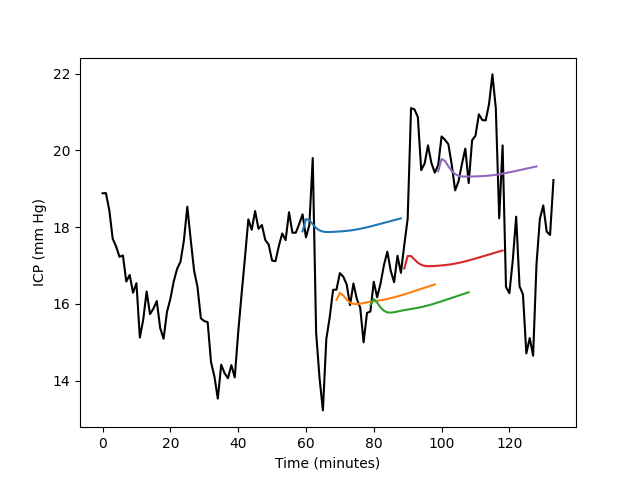}
    \caption{30-minute forecast every 10 min (LSTM).}
    \label{fig:pred_5_min_LSTM}
  \end{minipage}
\end{figure}

To illustrate how the predictions change over time, we zoom in on a specific (volatile) segment of 130 minutes. Here we make a new 30-minute prediction every 10 minutes, based on the past 60 minutes (Figure \ref{fig:pred_5_min_moment} and \ref{fig:pred_5_min_LSTM}). The black line indicates the observed signal and the colored lines indicate separate 30 minute predictions. For both the MOMENT and LSTM models, the the level of the predictions changes over time, while the shape of the line stays similar. The MOMENT model produces a more variable forecast (squiggly lines) than the LSTM model (straight lines). It is evident that both models lack the ability to predict high-magnitude, low-frequency changes in the observed signal.

\subsubsection{Performance by segment:}

\begin{figure}
    \centering
    \includegraphics[width=\linewidth]{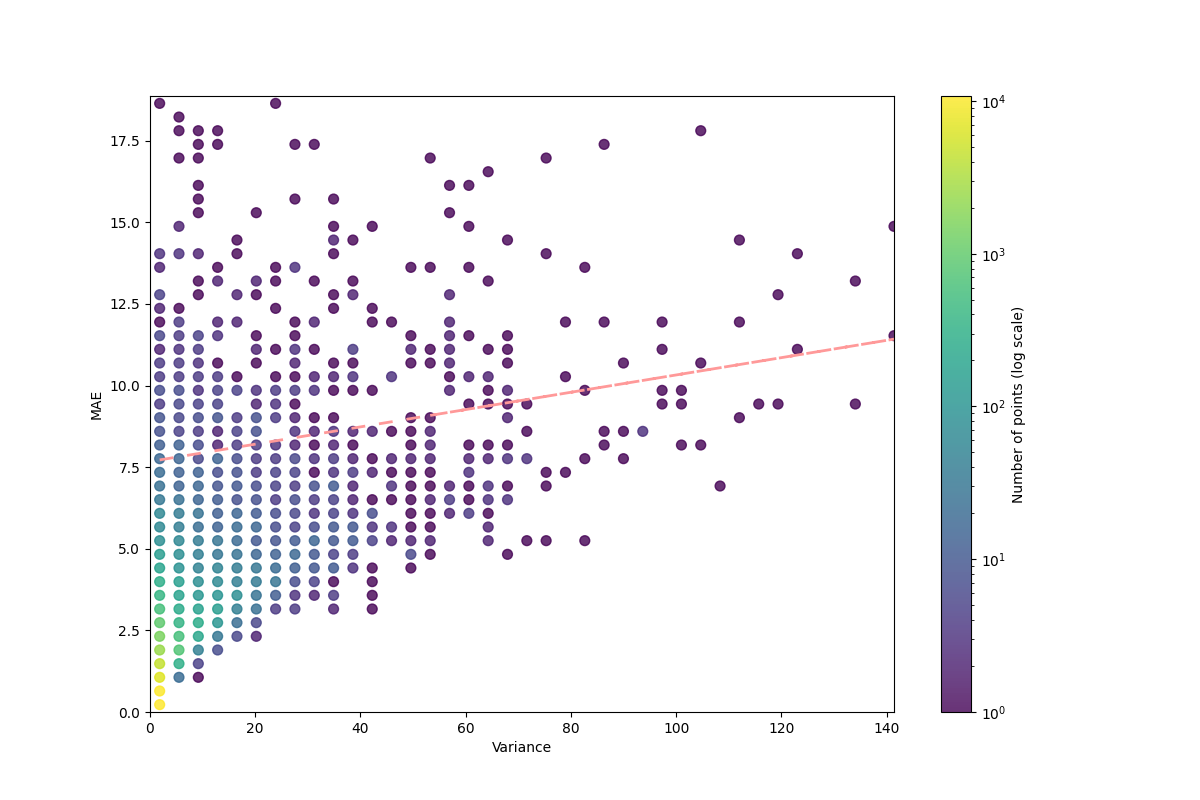}
    \caption{MAE vs variance for each segment (MOMENT).}
    \label{fig:change_moment}
\end{figure}

It seems that the models are not effective at forecasting changes in the signal, focusing instead on the general trend. We can demonstrate this by examining the relationship between performance (MAE) of the forecast in a segment (Equation \ref{eq:segment}) and the variance of the ICP itself within that observed segment. The variance in a segment reflects stability; if the variance is low, the signal remains relatively stable over time. Figure \ref{fig:change_moment} illustrates this relationship for the MOMENT model, showing all segments in the validation sets. The color of the points indicates the density of the segments, revealing that most points correspond to segments with low variance and low MAE. The red line represents a linear model depicting the relationship between variance and MAE. It appears that the more variable a signal is, the higher the MAE. Note that there are some points where the variance is low and the MAE is high, which can be explained by a sudden change in the signal followed by a stable period that is not predicted well by the model. The patterns observed for the MOMENT model are mirrored in the LSTM results (Figure \ref{fig:change_LSTM}, Appendix D).

\subsubsection{Performance by patient:}

\begin{figure}[h!]
    \centering
    \includegraphics[width=0.8\linewidth]{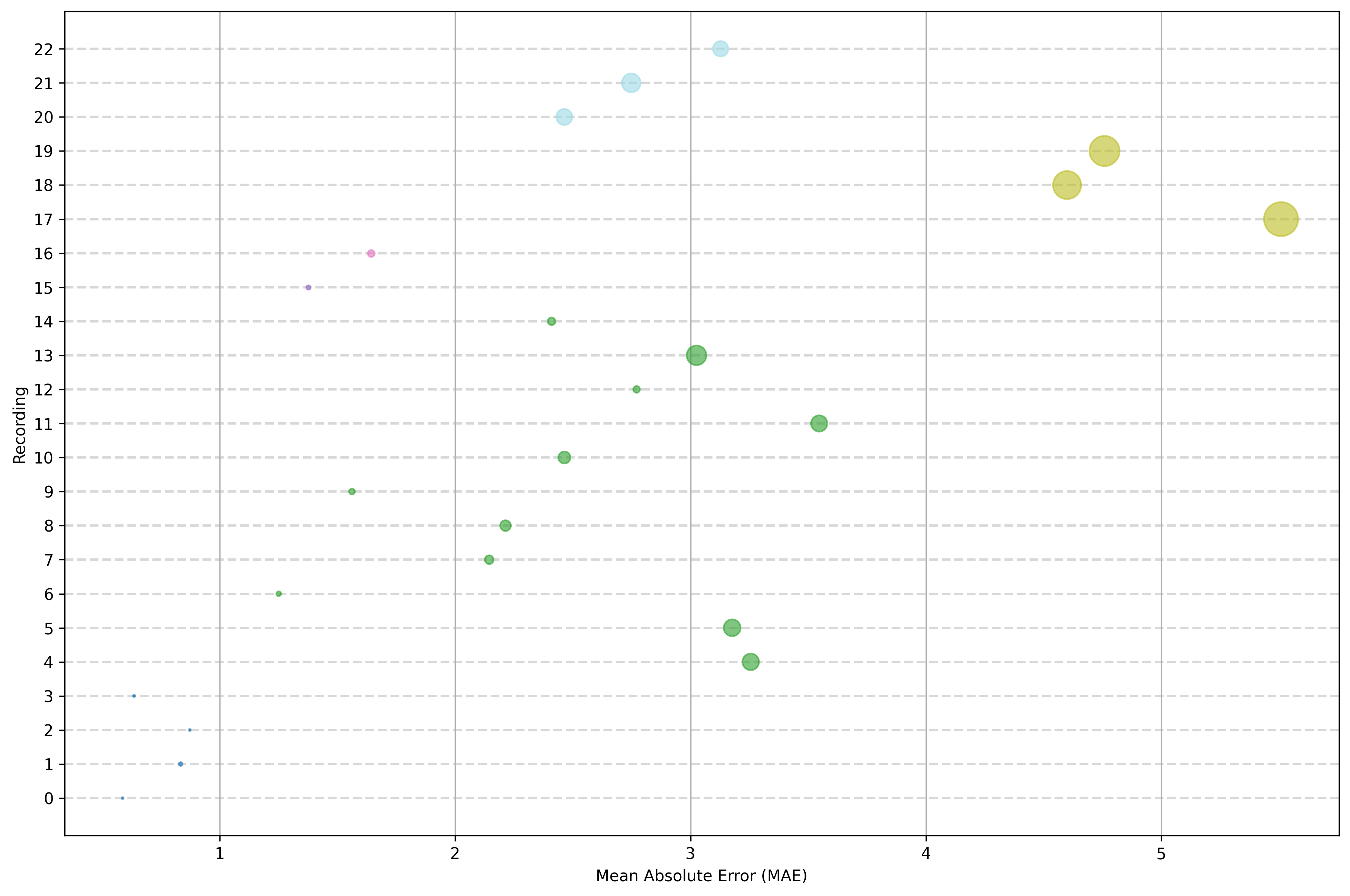}
    \caption{The MAE of the internal validation recordings from one CV run (MOMENT), grouped by patients (color of the points). The size of the points indicates the variance in the recording.}
    \label{fig:meta_mom}
\end{figure}

In Figure \ref{fig:meta_mom}, we observe the MAE for different validation patients using the MOMENT model. The size of the points indicates the average variance in the segments of a patient; a larger size corresponds to a greater average variance. Patients with multiple recordings are represented by multiple points with the same color. From the figure, we can make the following observations: there is substantial variance between patients, and recordings of a patient tend to cluster around the same MAE value. Additionally, larger points correspond to higher MAE, consistent with previous observations. A similar pattern is observed for the LSTM model, as shown in Figure \ref{fig:meta_lstm} (Appendix D).

\subsection{External validation}

The results from the external validation are shown in Table \ref{tab:validation_ye}. The results show that the MOMENT and the LSTM models perform very similarly and outperform the ES model. The scale of the performance is comparable with that of the internal validation (Table \ref{tab:model_comparison_val}). The training results are shown in Table \ref{tab:training_performance_all} (Appendix D). 



\begin{table}[htbp]
\caption{External validation performance.}
\centering
\label{tab:validation_ye}
\begin{tabular}{|c|c|c|c|}
\hline
\textbf{Metric} & \textbf{MOMENT} & \textbf{LSTM} & \textbf{ES} \\
\hline
MSE & 9.64 & 9.56 & 24.77 \\
MAE & 1.95 & 1.92 & 3.43 \\
90th percentile MAE & 4.13 & 4.02 & 7.03 \\
99th percentile MAE & 9.25 & 9.22 & 14.13 \\
\hline
\end{tabular}
\end{table}

 The MAE per segment versus variance per segment has a similar pattern as before (Figures \ref{fig:MAE_var_mom_ye} and \ref{fig:MAE_var_lstm_ye}, Appendix D). The upper bound for the variance in the segments is higher, indicating that there are more "difficult" segments in the CHARIS dataset compared to the TRACK-TBI dataset.

\begin{figure}[h!]
    \centering
    \includegraphics[width=0.75\linewidth]{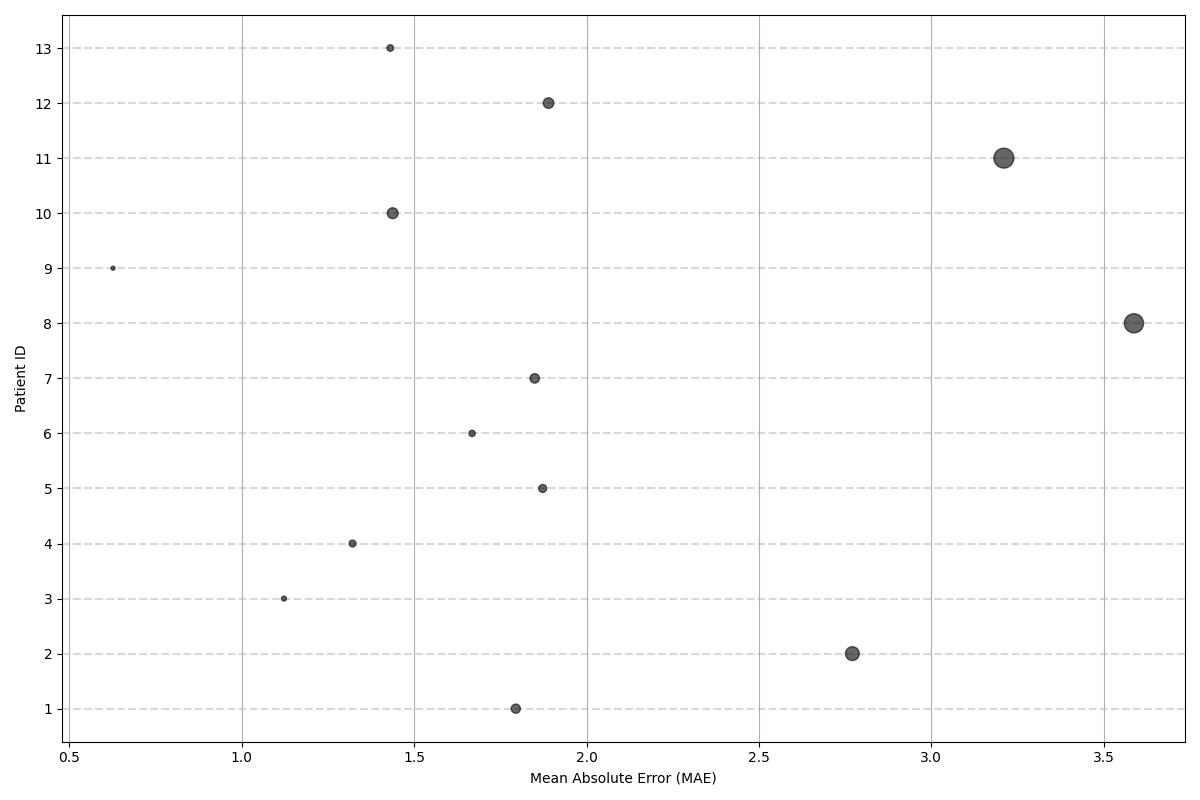}
    \caption{The MAE of the external validation patients (MOMENT). The size of the points indicates the variance in the recording.}
    \label{fig:meta_mom_ye}
\end{figure}

In Figure \ref{fig:meta_mom_ye}, we show the MAE for each patient in the external validation set for the MOMENT model.
In this dataset, each patient only has one recording. There is again a large spread in the observed MAE values. The larger points, with more variance in the signal, seem to have higher MAE then the smaller points. The LSTM model's performance characteristics closely resemble those of the MOMENT model (Figure \ref{fig:meta_LSTM_ye}, Appendix D).

\section{Discussion}
We developed the first fine-tuned foundation model for forecasting 30 minutes of ICP signal (MOMENT) and evaluated its performance against broadly used time series models (LSTM and ES). We measured their performance using internal validation (TRACK-TBI) and external validation (CHARIS). We have shown that when forecasting the ICP signal, based on the past 60 minutes of the signal, the models do not perform well. Variable segments cannot be predicted well, and there is a large variation in the performance between patients.

The results are worse than those reported in \cite{ye_machine_2022}, where an RMSE of 2.18 (MSE = 4.75) was achieved. We believe this discrepancy can be partly explained. Firstly, the choice was made to forecast 10 minutes ahead, while we forecast 30 minutes ahead. This highlights the principle that the further into the future we forecast, the more challenging it becomes to make accurate predictions. Secondly, with only 13 patients available, they opted to use only a training set in the development and assessment of their model. Additionally, the data they used come from a single site, which might have helped the model learn more effectively. In contrast, our training comes from three different sites, which may introduce more variability into the data. 

The models used in this study differ in their ease of use. We observed the following when using the different models. The ES model is very easy to use and converges quickly. The LSTM model trains quickly and can also be easily adapted to take in multiple features or to adjust the length of the input/forecast. The MOMENT model was most difficult to implement. The input window for MOMENT is fixed at a size of 512. This means that in many cases, zero padding needs to be used, which is computationally wasteful as the "real" input length is then shorter than the one the model uses to produce predictions. The training (fine-tuning) of the MOMENT model is slow compared to that of the LSTM; in this study, training the MOMENT model took about 10 times longer than the LSTM model.

A limitation of our approach is that we do not account for the uncertainty around our predictions, as a clinician might want to know how certain we are about a particular forecast before using it in practice. For the MOMENT model, the architecture is fixed, preventing us from using a Bayesian neural network \cite{izmailov_what_2021}, which would also be challenging due to computational constraints. Other options that could be used for obtaining uncertainty estimates include Monte Carlo dropout \cite{gal_dropout_2016} and the conformal prediction framework for time series \cite{stankeviciute_conformal_2021}.


Before a clinical decision support model could actually be implemented, we need to be sure it will work in different settings. The performance observed in this study is likely an underestimate of how well the model would work in an external setting, as evidenced by the discrepancy between the training and external validation performance. Therefore, before considering implementation, the model should be externally validated (multiple times) in similar settings to where they will be applied \cite{van_calster_there_2023}. If external validation occurs in many settings then one could create a prediction interval for how well the model will perform in a new setting \cite{van_leeuwen_empirical_2024}. This prediction interval could serve as a basis for a check to ensure there is not too much uncertainty to implement the model. If we see that the prediction interval is very wide, then it might be a good idea to fine-tune the model with some local data before implementation.

We conclude this work by noting that based on our research, ICP forecasting based solely on the signal does not achieve sufficient performance for practical implementation. The concept of incorporating prior information into a model has a strong theoretical advantage, which was only partially realized in this study. We speculate that the ICP signal may differ too significantly from the signals used in the training of the MOMENT model, or alternatively, that the unexplained variance in the ICP signal is simply too large.

\begin{credits}
\subsubsection{\ackname} We would like to thank Sebastian Mildiner Moraga for his comments on an initial version, Justin Moore and Tony Kabilan Okeke for their assistance in accessing the data, and Hruday Vairagade and Hiep Nguyen for valuable discussions during an open-air seminar.

\end{credits}

\newpage
%
%
%
\bibliographystyle{unsrt}
\bibliography{ICP_bib}
\newpage

\section*{Appendix A}

\begin{figure}[hbt!]
    \centering
    \includegraphics[width=0.7\linewidth]{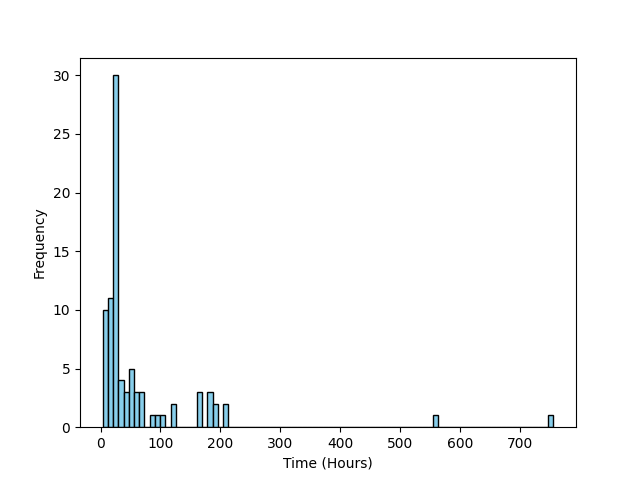}
    \caption{Frequency of the recording times of the ICP recordings from TRACK-TBI.}
    \label{fig:recording}
\end{figure}

We see in Figure \ref{fig:recording} that most recording sessions have a duration of less than 100 hours (approximately 4 days). There are a few recordings with a much longer recording time.

\newpage

\section*{Appendix B}

\begin{figure}[hbt!]
    \centering
    \includegraphics[width=.9\linewidth]{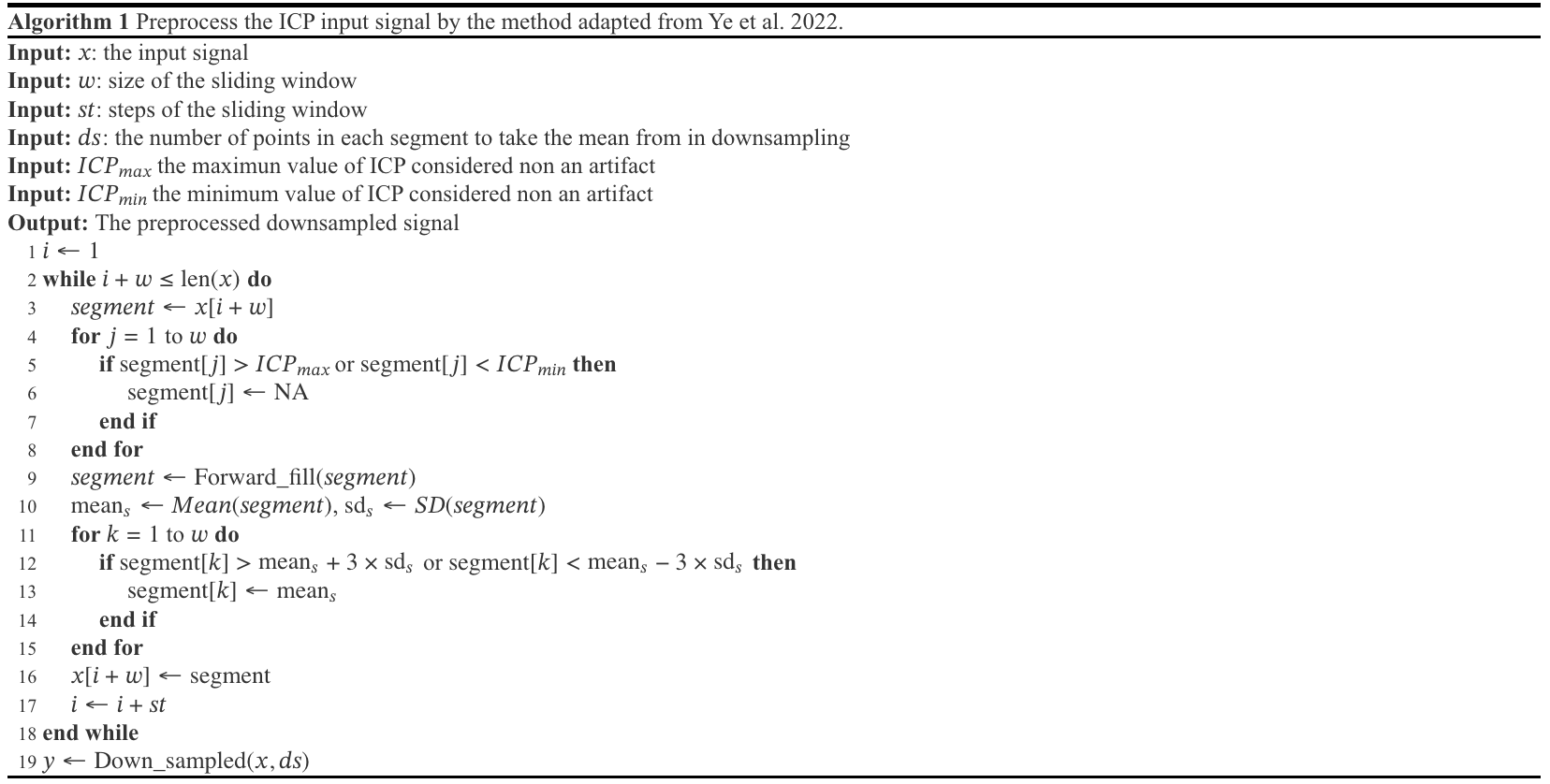}
    \caption{Preprocessing algorithm}
    \label{fig:pre_process}
\end{figure}

In Figure \ref{fig:pre_process} the preprocessing algorithm is outlined. The goal of the algorithm is to remove artifacts, smooth the time series, and downsample the data to make training computationally feasible. The input is a signal with 50 measurements every second (50 Hz). In the TRACK-TBI data not all recordings were measured at 50 Hz, so we look the mean for every 1/50 seconds. Some functions used in the algorithm are explained here:

\begin{itemize}
    \item Forward\_fill():  If a value is missing, the next available value is used to fill it in. 
    \item Down\_sampled(): The time series is downsampled to 1/ds of the original length by taking the mean. 
\end{itemize}

For the inputs in Figure \ref{fig:pre_process} we use the following values: $w$ = 60000 (20 minutes), $st$ = 3000, $ds$ = 3000, $ICP_{max}$ = 50, $ICP_{min}$ = -5 \footnote{The ICP values depend on the position of the patient, negative values can be observed when a patient is in an upright position \cite{norager_reference_2021}}.

\begin{figure}[hbt!]
    \centering
    \includegraphics[width=.9\linewidth]{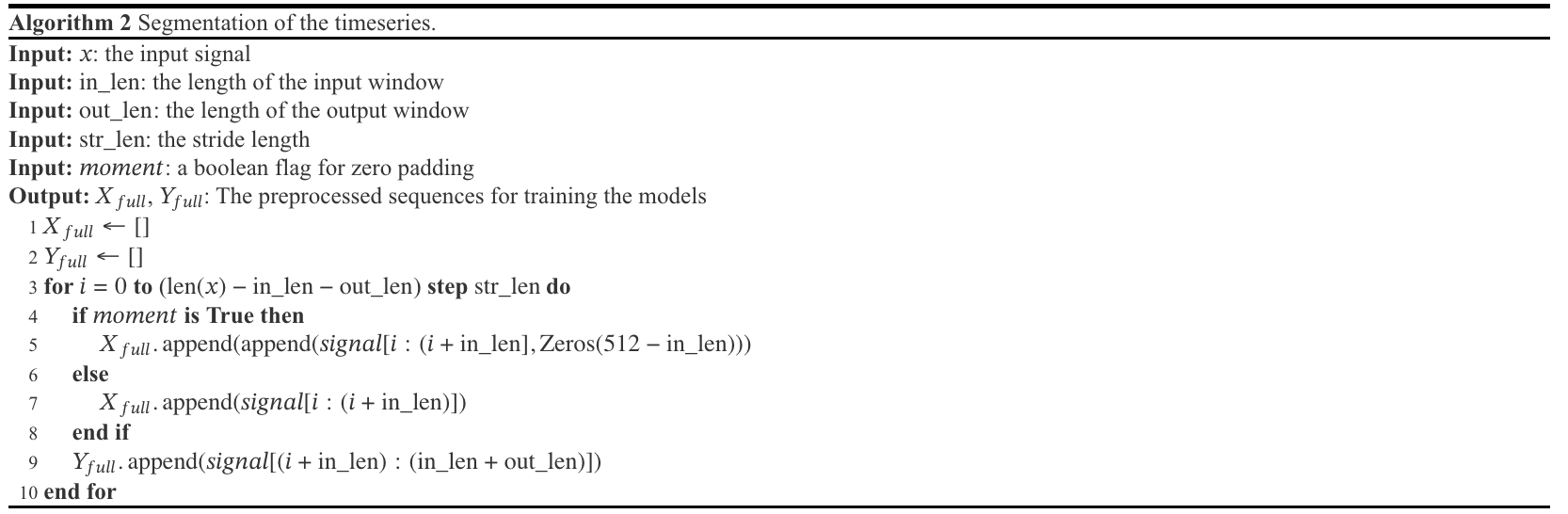}
    \caption{Segmentation algorithm}
    \label{fig:segmentation}
\end{figure}

Similarly, in Figure \ref{fig:segmentation}, the segmentation process is outlined. The goal is to transform a time series into smaller parts that can be used for training the model. We start by taking the first 'in\_len' data points from the time series as the first 'X', and the adjacent 'out\_len' data points as the 'Y'. We then move forward in the time series by a step size of 'str\_len', and perform the same procedure. This results in multiple 'X' and 'Y' segments from a single time series, with many segments containing overlapping data points. The 'Y' segments always follow the corresponding 'X' segments. The Zeros function creates a vector of length (512 - 'in\_len') to append to the segment. This is necessary for the MOMENT model, as the input vector has a fixed length of 512. For the inputs in Figure \ref{fig:segmentation}, we use the following values: in\_len = 60, out\_len = 30, str\_len = 5.

\begin{figure}[!tbp]
  \centering
  \begin{minipage}[b]{0.45\textwidth}
    \includegraphics[width=\textwidth]{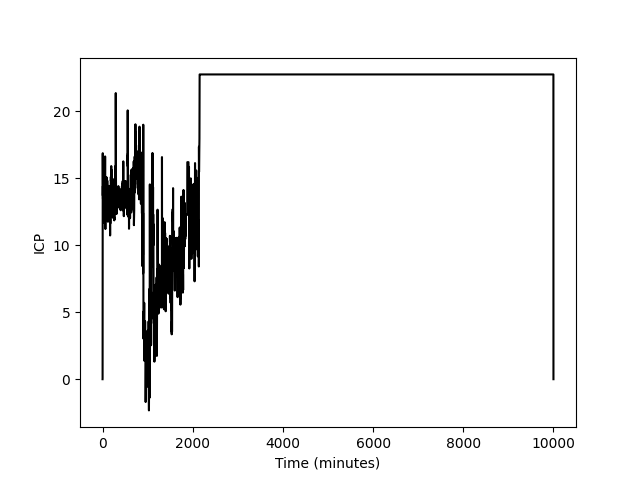}
    \caption{Unrealistic signal}
    \label{fig:pre_proc_bad_full}
  \end{minipage}
  \hfill
  \begin{minipage}[b]{0.45\textwidth}
    \includegraphics[width=\textwidth]{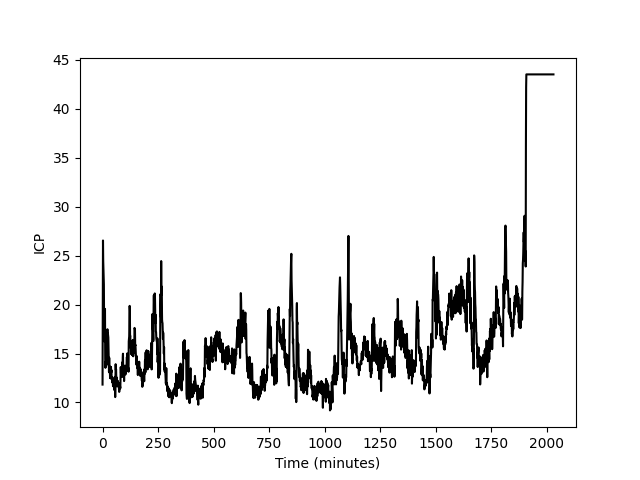}
    \caption{Unrealistic ending of signal}
    \label{fig:pre_proc_bad_end}
  \end{minipage}
\end{figure}

After the preprocessing, there were some signals that we considered unrealistic. An example can be seen in Figure \ref{fig:pre_proc_bad_full}. The straight section in the middle of the time series is caused by an excessive number of missing or unrealistic values, which were set to missing. We do not want the models to learn such behavior.

For some of the segments, we observed that the end of the signal appeared distorted. This could be due to disconnection during measurement. An example can be seen in Figure \ref{fig:pre_proc_bad_end}. For these signals (6 out of 83), we removed the last part of the signal.
 
\begin{figure}[]
    \centering
    \includegraphics[width=\linewidth]{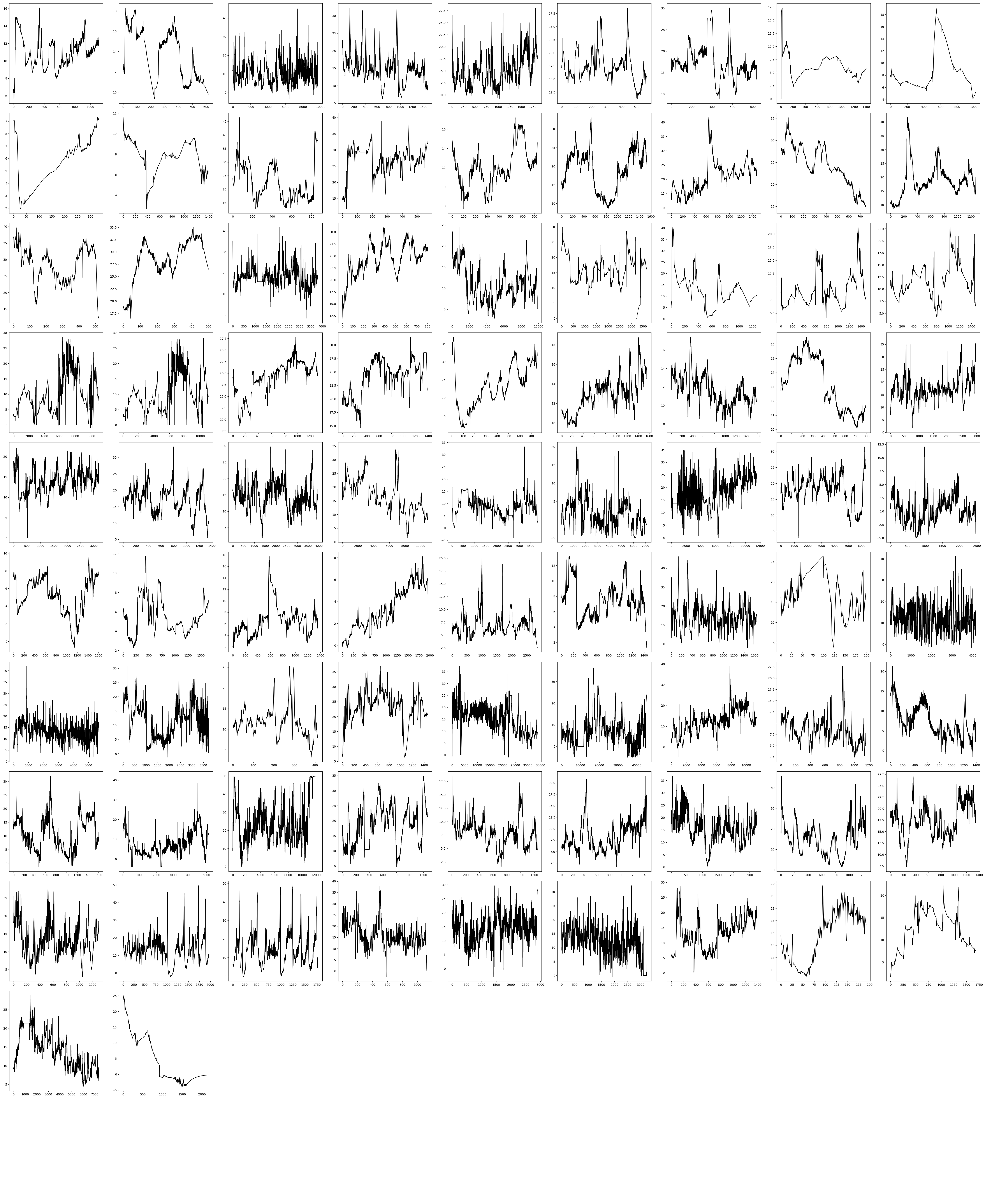}
    \caption{All the preprocessed signals from TRACK-TBI.}
    \label{fig:all_signals}
\end{figure}

In Figure \ref{fig:all_signals}, we see all the signals from TRACK-TBI used for training and evaluating the model.

\newpage

\section*{Appendix C}
The performance measurements can be denoted as follows. The MSE of a segment is defined as:
\begin{equation}
    MSE_{segment} = \frac{1}{s}\sum_{i=1}^s (y_i - \hat{y}_i)^2
\end{equation}

where $i$ represents the number of data points in a segment, $y_i$ the observed ICP value, $\hat{y}_i$ the predicted ICP value and $s$ is the total number of data points in the segment. Similarly for the MAE:

\begin{equation}
    MAE_{segment} = \frac{1}{s}\sum_{i=1}^s |y_i - \hat{y}_i|
    \label{eq:segment}
\end{equation}

To obtain the complete MAE of a patient, we need to take the average MAE over all segments:
\begin{equation}
    MAE_{patient} = \frac{1}{k} \sum_{j=1}^k MAE_{segment_j}
    \label{eq:patient}
\end{equation}

with $k$ being the number of segments. The overall MAE is then the average over all patients:
\begin{equation}
    MAE_{model} = \frac{1}{N}\sum_{n=1}^N MAE_{patient_n}
    \label{eq:total}
\end{equation}

With $N$ being the number of patients. 

\newpage

\section*{Appendix D}

\begin{figure}[hbt!]
  \centering
  \begin{minipage}[b]{0.49\textwidth}
    \includegraphics[width=\textwidth]{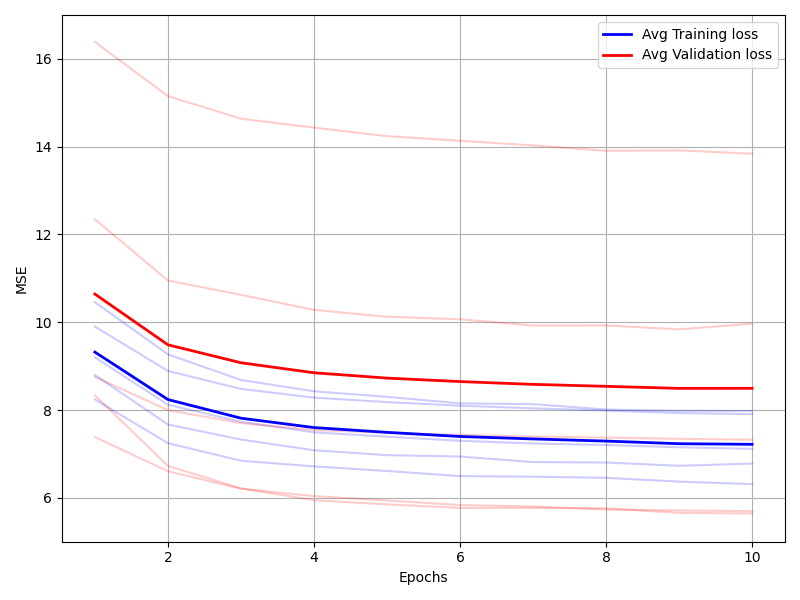}
    \caption{Loss vs epoch for MOMENT.}
    \label{fig:loss_moment}
  \end{minipage}
  \hfill
  \begin{minipage}[b]{0.49\textwidth}
    \includegraphics[width=\textwidth]{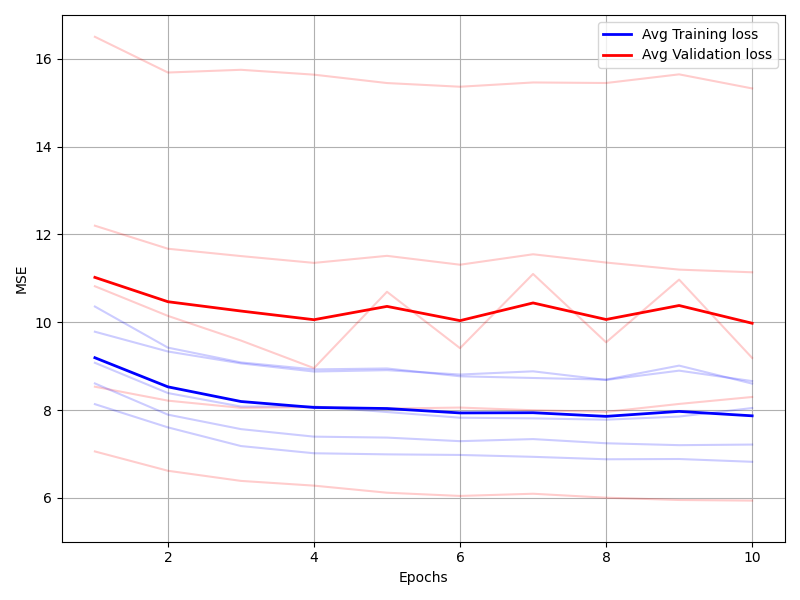}
    \caption{Loss vs epoch for LSTM.}
    \label{fig:loss_lstm}
  \end{minipage}
\end{figure}

For the internal validation, the training/validation loss per epoch can be seen in Figure \ref{fig:loss_moment} and \ref{fig:loss_lstm} for the MOMENT and LSTM model respectively. The bright lines are the mean losses over the CV runs, and the shaded lines are the CV runs themselves. There are two validation runs with a very high loss; after inspection of the signals, it seems that in both validation sets there was one patient with multiple recordings with a very volatile signal. The MOMENT model seems to converge quicker and more smoothly than the LSTM.

The training performance is a bit better than the validation performance (Table \ref{tab:training_performance}). Interestingly enough, the LSTM (slightly) outperforms the MOMENT model in the training set. It thus seems like the LSTM overfits on the training set, as the internal validation performance is better for the MOMENT model Table \ref{tab:model_comparison_val}.

\begin{table}[hbt!]
\caption{Average training performance over 5 CV folds, SD is in brackets.}
\centering
\label{tab:training_performance}
\begin{tabular}{|c|c|c|c|}
\hline
\textbf{Metric} & \textbf{MOMENT} & \textbf{LSTM} & \textbf{ES} \\
\hline
MSE & 7.77 (0.87) & 7.92 (0.89) & 20.22 (1.61) \\
MAE & 1.62 (0.11) & 1.61 (0.14) & 2.83 (0.13) \\
90th percentile MAE & 3.66 (0.23) & 3.60 (0.23) & 6.11 (0.29) \\
99th percentile MAE & 8.88 (0.43) & 9.03 (0.44) & 14.68 (0.32) \\
\hline
\end{tabular}
\end{table}

\begin{figure}[hbt!]
    \centering
    \includegraphics[width=\linewidth]{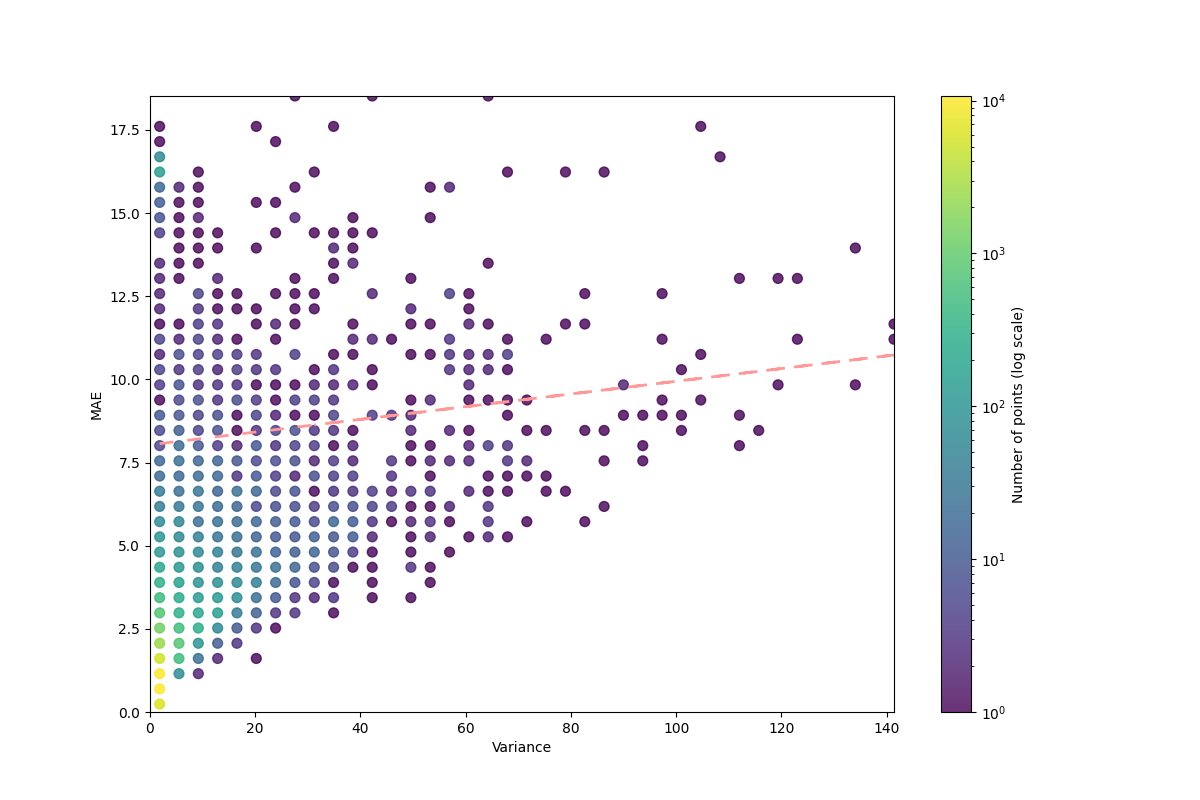}
    \caption{MAE vs variance for each segment (LSTM model).}
    \label{fig:change_LSTM}
\end{figure}

The MAE versus the variance per segment in the LSTM model is shown in Figure \ref{fig:change_LSTM}. It shows the same relationship as was seen in Figure \ref{fig:change_moment}; most segments have a low MAE and a low variance, and it seems that as the variance increases, the MAE increases. 

\begin{figure}[hbt!]
    \centering
    \includegraphics[width=\linewidth]{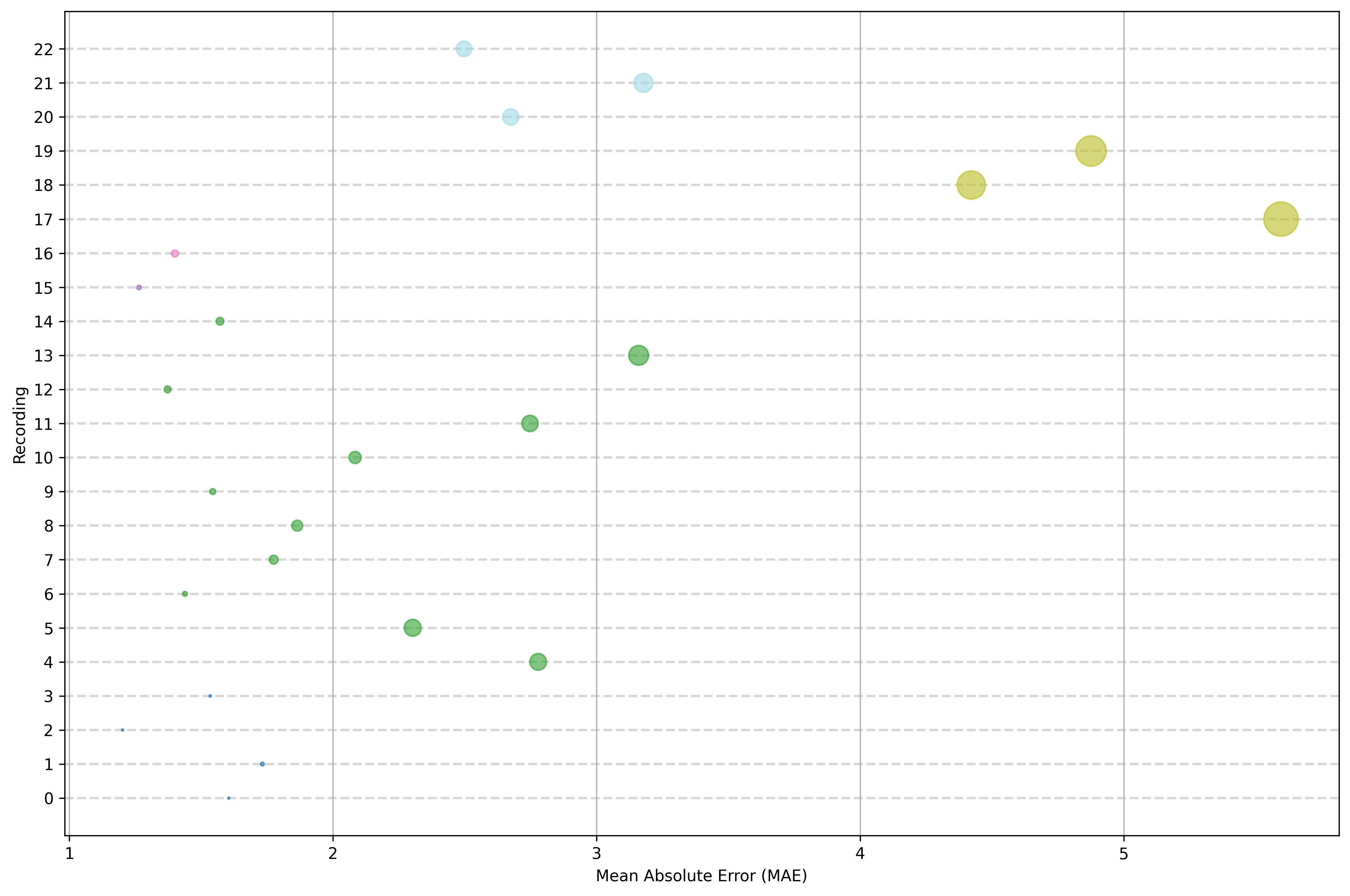}
    \caption{The MAE of the validation recording from one CV run (LSTM), grouped by patients (color of the points). The size of the points indicates the variance in the recording.}
    \label{fig:meta_lstm}
\end{figure}

The MAE per recording, grouped by patients, is shown in Figure \ref{fig:meta_lstm}. This is for the same CV run, but this time for the LSTM model. We again see a very similar pattern to Figure \ref{fig:meta_mom}.

The training performance when the model is trained using all available data is shown in Table \ref{tab:training_performance_all}. The performance is a bit better than the average training performance of the 5-fold CV (Table \ref{tab:training_performance}). The MOMENT and LSTM models still perform very similarly, and a lot better than the ES model. 

\begin{table}[hbt!]
\caption{Training performance using all TRACK-TBI data.}
\centering
\label{tab:training_performance_all}
\begin{tabular}{|c|c|c|c|}
\hline
\textbf{Metric} & \textbf{MOMENT} & \textbf{LSTM} & \textbf{ES} \\
\hline
MSE & 7.57 & 7.70 & 20.07 \\
MAE & 1.60 & 1.61 & 2.82 \\
90th percentile MAE & 3.61 & 3.58 & 6.08 \\
99th percentile MAE & 8.71 & 8.82 & 14.66 \\
\hline
\end{tabular}
\end{table}

\begin{figure}[h!]
  \centering
  \begin{minipage}[b]{0.49\textwidth}
    \includegraphics[width=\textwidth]{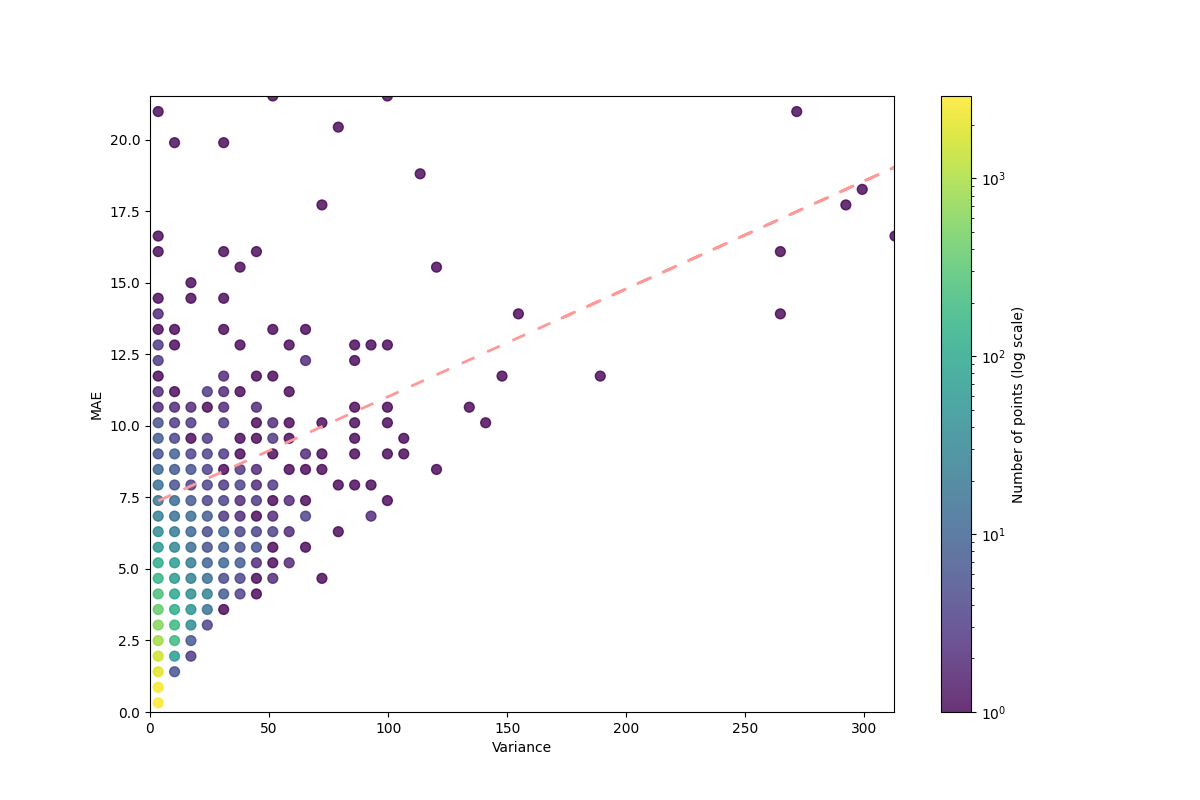}
    \caption{MAE vs variance (MOMENT).}
    \label{fig:MAE_var_mom_ye}
  \end{minipage}
  \hfill
  \begin{minipage}[b]{0.49\textwidth}
    \includegraphics[width=\textwidth]{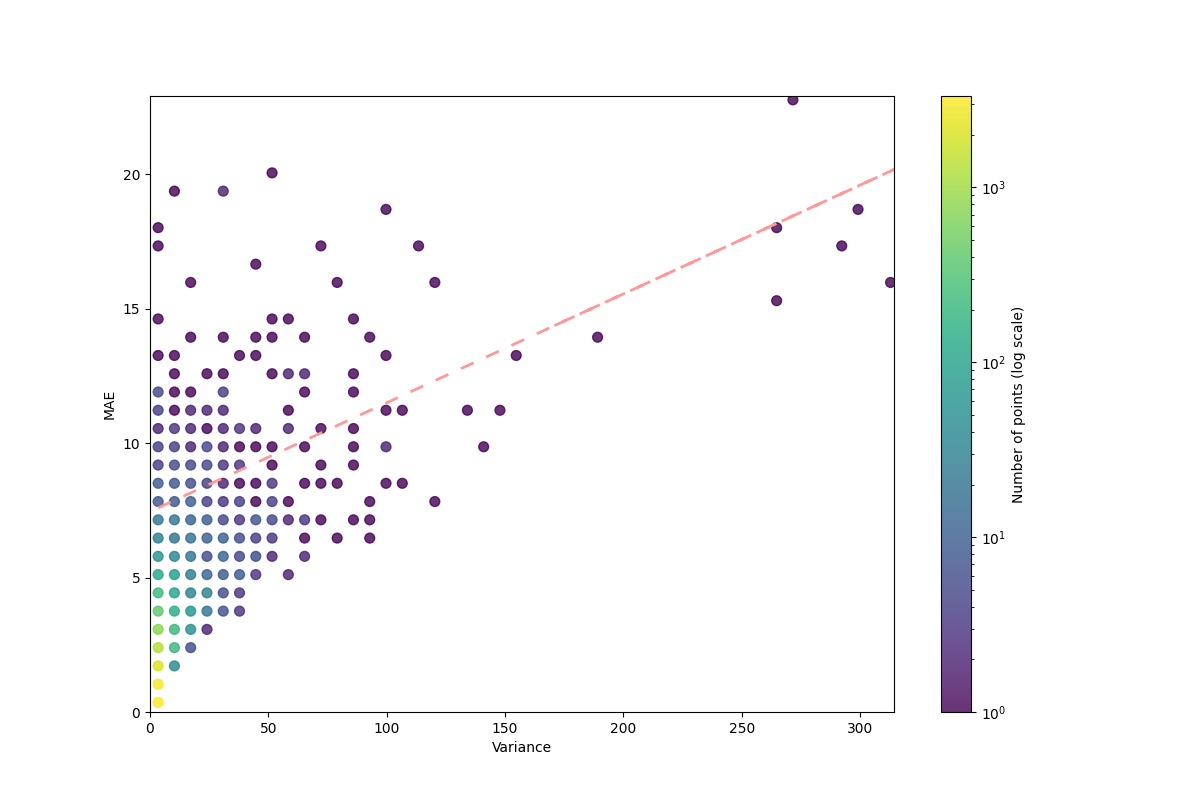}
    \caption{MAE vs variance (LSTM).}
    \label{fig:MAE_var_lstm_ye}
  \end{minipage}
\end{figure}

For the external validation dataset, we can also inspect the relationship between the variance in the segments and the MAE for the MOMENT and LSTM models (Figures \ref{fig:MAE_var_mom_ye} and \ref{fig:MAE_var_lstm_ye}). There seem to be segments with higher variance in this dataset compared to the TRACK-TBI data. The relationship between the MAE and variance is similar for the MOMENT and LSTM models and seems more extreme than in the internal validation (Figure \ref{fig:change_moment} and\ref{fig:change_LSTM}).

\begin{figure}[h!]
    \centering
    \includegraphics[width=\linewidth]{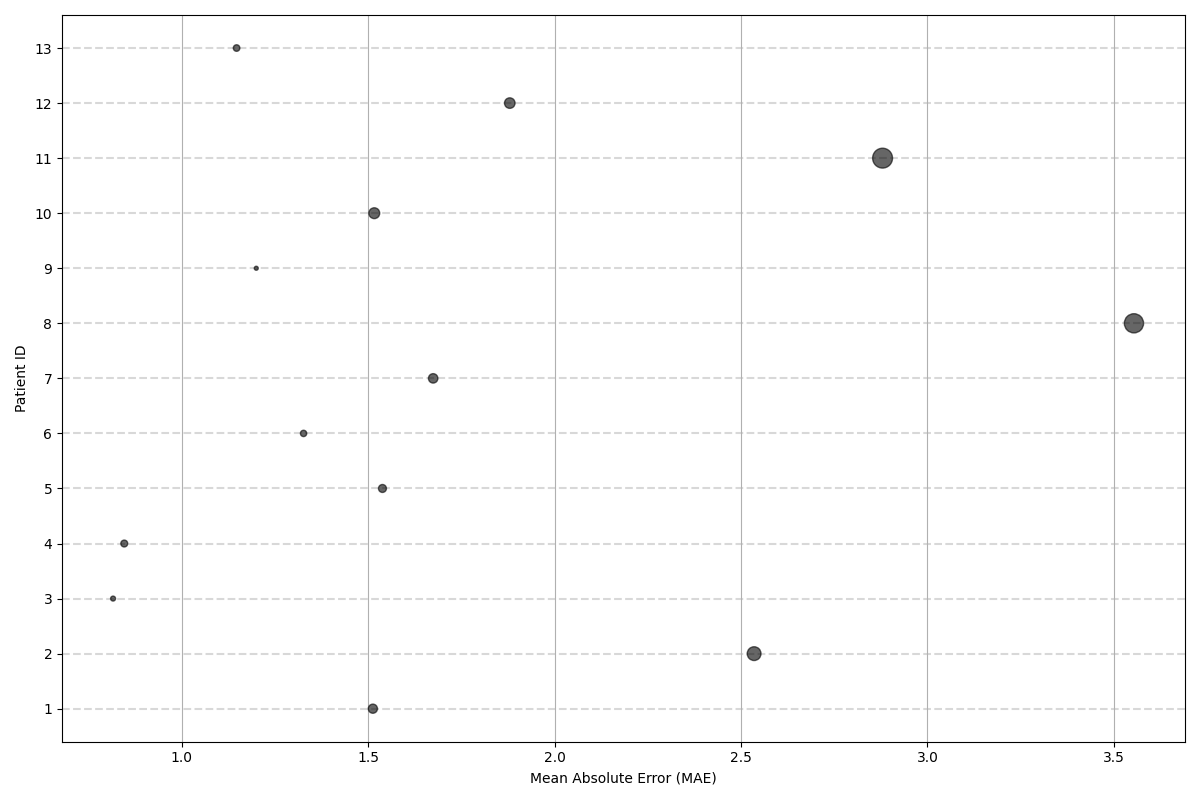}
    \caption{The MAE of the external validation patients (LSTM). The size of the points indicates the variance in the recording.}
    \label{fig:meta_LSTM_ye}
\end{figure}

The MAE of the LSTM versus the variance of in the signal of the external validations patients is shown in \ref{fig:meta_LSTM_ye}. 

\end{document}